# The SiPM Array Data Acquisition Algorithm Applied to the GECAM Satellite Payload

Y.Q. Liu[*,1], K. Gong[*,1], X.Q. Li[*,1], X.Y. Wen[*,1], Z.H. An[1], C. Cai[1,2], Z. Chang[1], G. Chen[1], C. Chen[1,2], Y.Y. Du[1], M. Gao[1], R. Gao[1], D.Y. Guo[1], J.J. He[1], D.J. Hou[1], Y.G. Li[1], C.Y. Li[1,3], G. Li[1], L. Li[1], X.F. Li[1], M.S. Li[1], X.H. Liang[1], X.J. Liu[1], F.J. Lu[1], H. Lu[1], B. Meng[1], W.X. Peng[1], F. Shi[1], X.L. Sun[1], H. Wang[1], J.Z. Wang[1], Y.S. Wang[1], H.Z. Wang[1], X. Wen[1], S. Xiao[1,2], S.L. Xiong[1], Y.B. Xu[1], Y.P. Xu[1], S. Yang[1], J.W. Yang[1], Q.B. Yi[1], Fan. Zhang[1], D.L. Zhang[1], S.N. Zhang[1], C.Y. Zhang[1], C.M. Zhang[1], Fei Zhang[1], X.Y. Zhao[1], Y. Zhao[1], X. Zhou[1]

[1]Key Laboratory of Particle Astrophysics, Institute of High Energy Physics, Chinese Academy of Sciences, Beijing 100049, China

[2]University of Chinese Academy of Sciences, Chinese Academy of Sciences, Beijing 100049, China

[3]Physics and Space Science College, China West Normal University, Nanchong 637002, China

[Abstract]: The Gravitational Wave Burst High-energy Electromagnetic Counterpart All-sky Monitor (GECAM), consists of 2 small satellites that each contain 25 $LaBr_3$ (lanthanum bromide doped with cerium chloride) detectors and 8 plastic scintillator detectors. The detector signals are read out using a silicon photomultiplier (SiPM) array. In this study, an acquisition algorithm for in-orbit real-time SiPM array data is designed and implemented, and the output event packet is defined. Finally, the algorithm's efficacy for event acquisition is verified.

[Keywords]: silicon photomultiplier, digital waveform processing, data acquisition

# 1 Introduction

Data acquisition is an essential part of space exploration. Real-time acquisition algorithms for astronomical satellites must meet the demands of space astronomy for determining the time and energy characteristics of signals under the space applications' numerous resource constraints. Several satellites have been launched to detect gamma-ray bursts (GRBs), some of which use the traditional single-channel triggering and peak-hold technique for data acquisition, and some of which use digital pulse waveforms for fast data sampling and online data processing.

The Shenzhou-2 spacecraft that was launched in 2001 carried a Gamma-ray Burst (GRB) monitor system (comprised of two thallium-activated sodium iodide (NaI(Tl)) scintillation detectors and one bismuth germanate (BGO, $Bi_4Ge_3O_{12}$) scintillation detector). The system acquired information using the typical single-channel triggering and peak-hold technique, followed by an analog-to-digital (AD) conversion; a microcontroller unit (80C196) was used for energy spectrum acquisition, count recording, and mode switching [1]. Field-programmable gate arrays (FPGAs) were subsequently used in many scientific payloads in aerospace applications due to their light weight and parallel logic. The gamma-ray burst monitor (GBM) aboard the Fermi satellite launched in 2018 contained 12 NaI(Tl) scintillation detectors and 2 BGO scintillation detectors. The data processing unit (DPU) of the GBM used an FPGA for digital signal processing, and each detector independently sampled signals using a 9.6-MHz analog-to-digital converter (ADC) fast-pulse waveform, with configurable dynamic digital thresholds and an average dead time of 2.6 μs. The GBM could output 4096 linear energy channels. The GBM possesses a considerably enhanced energy measurement capability and flexibility [2]. The space radiation detector onboard the SJ-10 Satellite launched in 2016 also applied the single-channel triggering and peak-hold technique, followed by an AD conversion, to its three detectors (1 silicon (Si) detector and 2 cadmium zinc telluride (CZT) detectors) and used an FPGA for data acquisition. The overall design was compact and low in power consumption, and the space radiation detector could complete online particle identifications and download particle energy channel results [3].

The payload of the Gravitational Wave-burst High-energy Electromagnetic Counterpart All-sky Monitor (GECAM) satellite uses different silicon photomultiplier (SiPM) arrays to match the detector crystals, which reduce the uniformity of light collection and resist magnetic field interference while measuring energy spectra and light variations, to localize the GRBs in orbit. We used an FPGA to perform digital pulse waveform processing and integrated the results into an event packet containing time and energy information. We then formed parallel channel data by using a data management FPGA to generate in-orbit trigger information using an in-orbit positioning algorithm. In this paper, we present the event discrimination and event packet processing methods for data acquisition. The results provide a reference for designers in their respective fields.

# 2 GECAM Satellites

The GECAM [4] was launched in 2020 and was comprised of 2 satellites for all-sky monitoring. Each satellite has a high sensitivity and a wide energy measurement range. The 25 gamma-ray detectors (GRDs) and 8 charged particle detectors (CPDs) identify particle bursts by coincidence measurements and performs time-varying and energy measurements of the GRBs.

Each GRD collects light through an array of 64 SiPMs (MicroFJ-60035-TSV). The array output signal is amplified into 2 different gain signals with tested energy ranges of 4.3 keV-300 keV and 40 keV-4.3 MeV. The CPDs are aggregated into 1 readout signal by an array of 36 SiPMs. The 33 detectors of each satellite output a total of 58 SiPM signals. The 58 signals are divided into 5 groups and are input to 5 data acquisition (DAQ) boards for event collection. As shown in Figure 1, the event information obtained from the 5 DAQs is processed and aggregated in 1 data management board (DM) before being transmitted to the satellite.

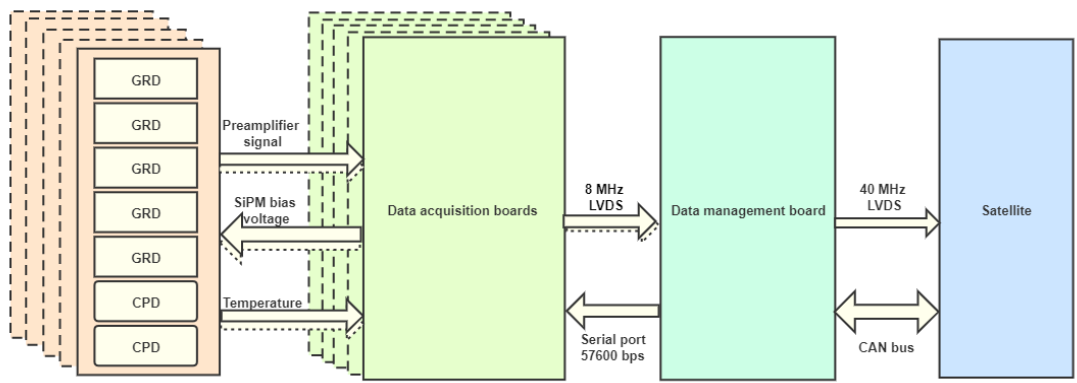

Figure 1 Schematic of the Electronic Connection of the GECAM Payload

Compared with the nonaerospace FPGA-based DAQ system, the GECAM DAQ board also needs to control and monitor the normal operations of the test. The DAQ board configures the detection activities according to the received instructions and processes the project operating parameters, such as voltage, current, flux and baseline data. Furthermore, the DAQ board collects the SiPM temperature and controls the overvoltage to stabilize the SiPM gain in response to temperature changes.

The 5 DAQ boards are independent of each other. The 12 different device interfaces of a DAQ board are controlled by one FPGA chip (Microsemi's SmartFusion2 FPGA - M2S090TS). The SmartFusion2 FPGA with a SEU immune Zero FIT Flash FPGA configuration is suitable for aerospace. The choice of M2S090TS-1FG484 is based on the reliability requirements of less than 80% resource usage, energy consumption restrictions and input/output (I/O) expansion.

# 3 Data Acquisition Design

We used a digital waveform processing algorithm for data acquisition under current resource constraints, and the output waveform of each detector is converted to a digital signal by a 40 MHz fast ADC by a DAQ board (Figure 2), and the digital signal is processed to obtain the linear 4096-channel energy information. As58 channels are grouped into 5 DAQ boards, and up to 12 data processing channels are needed for each DAQ board. All data packets in each DAQ board are downloaded via an 8-MHz low-voltage differential signaling (LVDS) interface.

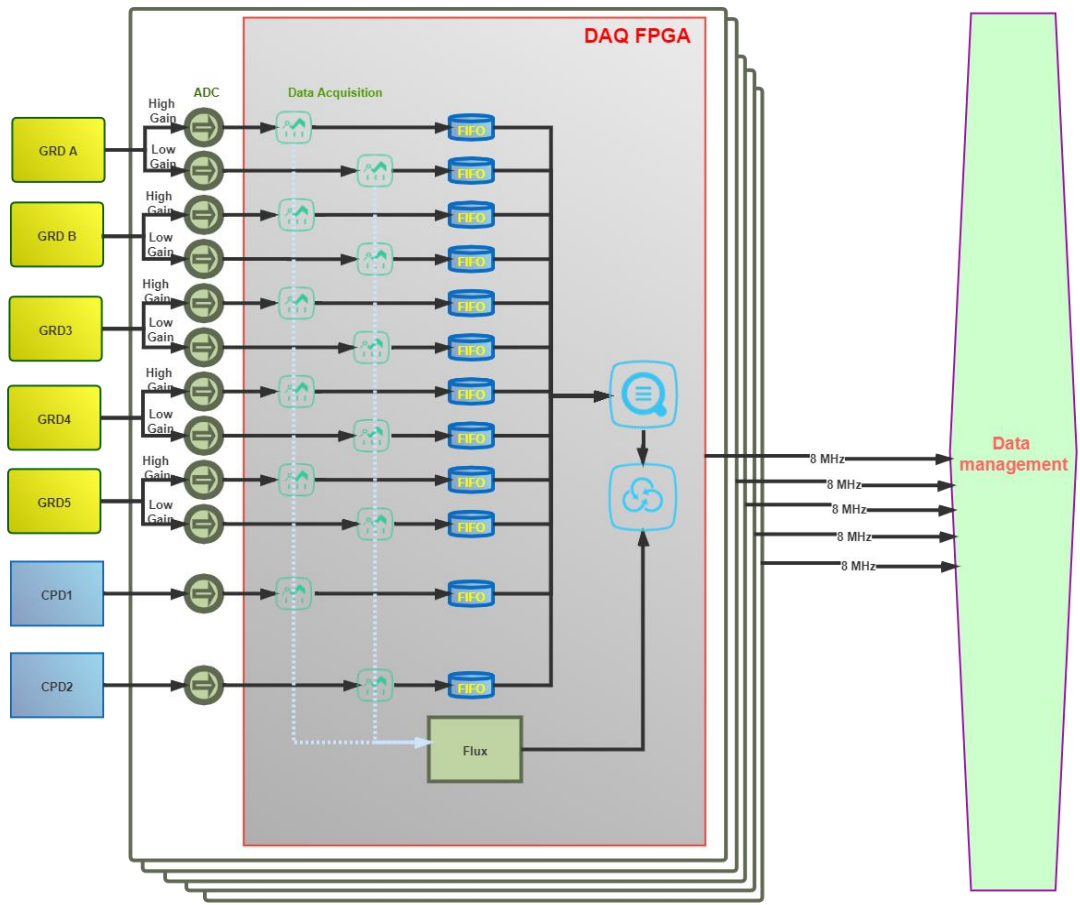

Figure 2 Block Diagram of Detector Signal Acquisition

The waveform processing and storage for 12 channels are parallel and independent. Each channel has a first-in-first-out (FIFO) cache for event packets. The size of each channel's FIFO depends on the in-orbit simulation. The FIFO cache will alleviate event packet loss caused by an extremely high count rate. To avoid data congestion caused by the huge flux of some channels, we also adopted "forced polling" of each channel's event packets before transmission. Therefore, the maximum flux of all channels can reach up to approximately 710 000 counts per second, which can handle most GRBs.

Additionally, the DAQ FPGA also generates two other types of data packets, engineering event packets and time event packets, and they have the highest sending priority. The engineering event packets record the flux (count) of each channel and other current and voltage telemetry data. The time event packets contain pulse per second (PPS) events, universal coordinated time (UTC) events and carryover time events.

## 3.1 SiPM Event Identification

The SiPM array of the GRDs and CPDs outputs an analog signal. Signal-conditioning amplifiers were used to make analog signals compatible with the voltage range of the fast ADC. The signal is directly converted into a digital signal via a fast ADC before entering the FPGA.

The digital signal processing algorithm is illustrated in Fig. 3. The digital signal is input into the baseline calculation and SiPM event identification module at the same time. The average value of digital signals is generated in real time to help identify SiPM events. SiPM event identification finds the impulse response of the detector and outputs the trigger. When the trigger is valid, the pulse height analyzer module obtains the peak of the pulse as energy. A local time counter is set in the FPGA. The local time corresponding to the trigger moment is recorded as the arrival time of the signal. The trigger duration count is also recorded synchronously. If the duration exceeds the normal dead time, this case will be marked. In the event classification module, according to the tag information, the time information and energy information of the event are packaged in different ways.

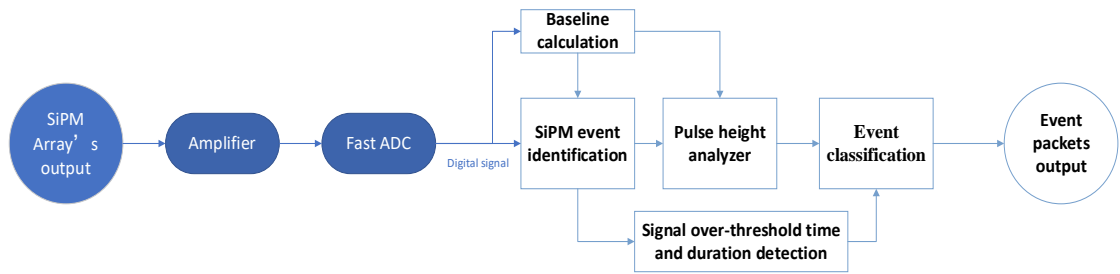

Figure 3 A functional overview showing the details of the digital signal processing algorithm

A timing diagram of the waveform model and trigger generator that are used for signals with different gains are shown in Figure 4. The rise time of the signal is approximately 300 ns, the pulse over the threshold duration is approximately 2 μs, and the end time of the signal undershoot is approximately 4 μs. Due to the high gain of the GRD signal, an oscillation occurs when the signal returns to the baseline. It is possible that a false trigger is generated when the falling edge of the digital signal is jittered. We use two threshold settings to build a digital Schmitt trigger to reduce the false trigger. When the rising data of the signal exceeds the high threshold, the trigger is valid; only when the falling data are lower than the low threshold does the trigger become invalid.

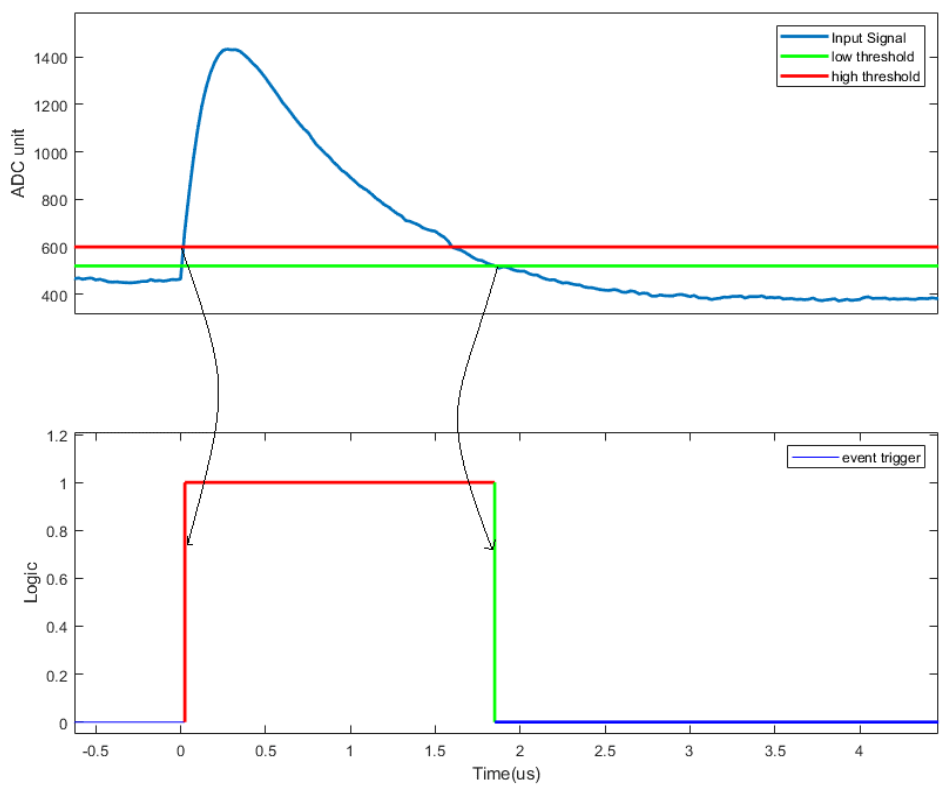

Figure 4 Timing diagram of event identification. The upper figure shows a waveform of the SiPM output signal after digital sampling. The red and green lines are the thresholds set for the trigger. The figure below shows the trigger in relation to the two thresholds. The start time of the trigger (red vertical line) is recorded as the arrival time of the signal in each event packet. The duration of the trigger (the horizontal interval of the red and green lines) is also recorded as the dead time, which affects the event classification.

Two different acquisition modes are set to provide flexibility for treating signals with different gains, the fixed threshold acquisition and baseline-deducted threshold acquisition. The fixed threshold acquisition uses variable threshold values, but the threshold values are fixed at the beginning and end of the signal. The baseline-deducted threshold acquisition determines whether the signal is rising or falling based on the difference between the digital value after pulse conversion and the baseline. If the difference is higher than the high threshold, the signal is considered rising; conversely, if the difference is lower than the low threshold, the signal is considered falling.

In respect to the rising and falling characteristics of the signal, a continuous rise of more than 100 ns is considered the beginning of an event, and four consecutive falls of more than 100 ns are considered the end of an event.

## 3.2 Types of Event Packet

Event packets containing energy information are mainly classified according to the energy and duration of the events, as shown in Table 1.

Table 1 Event Classification

|  | ADC<x"FFF" | ADC=x"FFF" |
|---|---|---|
| Duration < set normal signal dead time (4 μs) | Normal signals: Download energy value | Saturation signal: delay of approximately 70 μs Download processing time |
| Duration ≥set normal signal dead time (4 μs) | Ultrawide signal: Download processing time | Saturation signal: delay of approximately 70 μs Download processing time |

The duration is mainly compared with the adjustable normal signal dead time (the default value is 4 μs, which is the undershoot time of the signal).

A normal event packet contains only energy information and no dead time information, and the default dead time for each normal event is the normal dead time. The signal is not cached if the interval between a signal and its preceding signal is less than the normal event dead time.

When the event duration of an unsaturated signal exceeds the normal dead time, the signal is recorded as an ultrawide signal. For an ultrawide signal, the compressed energy value is recorded, and the event duration is cached.

For signals with saturated sampling values, the ultrahigh dead time must be delayed before judging whether the signal ends. The ultrahigh dead time is adjustable with a default value of 70 μs to eliminate false signals induced by the overshoot of the saturated signals. For saturated signals, saturation identification is recorded, and the overall processing time is stored.

The event packets that include time information mainly contain global position system (GPS) events, UTC events, and carryover events.

Each data acquisition board has its own independent crystal oscillator (40 MHz) to perform local time counting (the minimum counting accuracy is 0.1 μs to reduce storage requirements) and generate the local time code. The local time code consists of two parts, a 24-bit high time code with the metric STime, and a 24-bit low time code with the metric PTime. The PTime count fills every 1.6777215 s and then carries over to STime to generate a carryover time packet. The absolute time of arrival (GNSS pulse) is recorded as a local time code to create a PPS packet, whereas the UTC time corresponding to the GPS time is received through a broadcast to generate a UTC packet. The time-information packets can be decoded to obtain the absolute time of arrival of each event.

## 4 Test Results

### 4.1 Introduction to overall performance

The lower the energy of GRB, the more astronomical information it contains. The energy range proposed by GECAM is 8 keV, which is lower than the current similar on-orbit instruments. We used an Fe-55 radioactive source at -20 ℃ to test the low energy threshold. The test result (Figure 5) has a low energy threshold of 4.3 keV, which meets the requirement. More results will be recorded in an upcoming dedicated paper. (Gong et al. Submitted)**.**

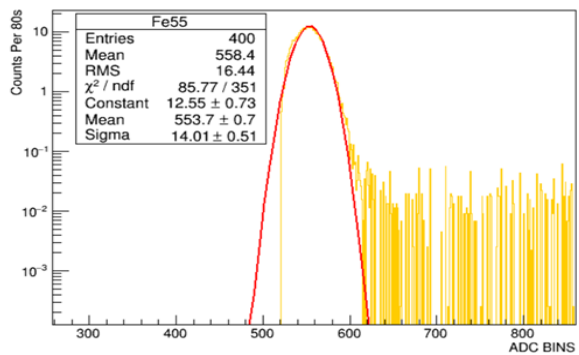

Figure 5 The test energy spectrum of Fe55; the red curve is the fitted curve. Yellow is the 5.9 keV peak after deducting the background of LaBr$_3$ and the noise

To verify whether the dead time is recorded accurately, the data of 2 channels of the same GRD are used for a cross-comparison. We used Cs137 and Am241 radioactive sources. The energies of 662 keV and 59.5 keV produce different types of event packets in two different gain channels. The test energy spectra are calibrated according to the dead time. As shown in Figure 6, the two intersecting energy ranges can be in good agreement. This shows the accuracy of the dead time records.

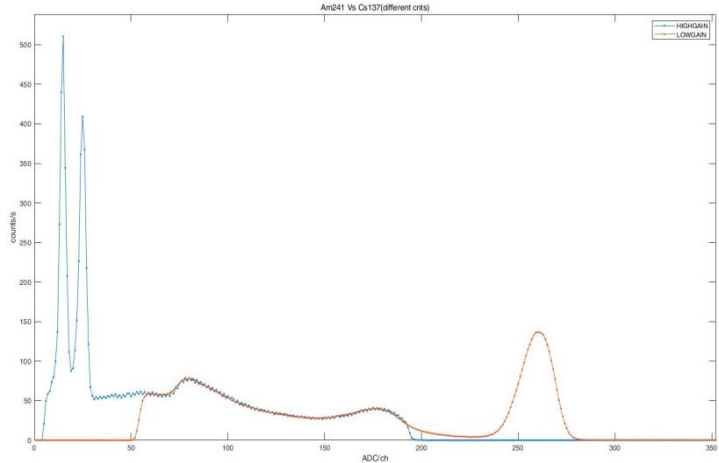

Figure 6 Comparison chart of the high- and low-gain energy spectra of the same GRD detector after dead time correction

To conduct joint observations with other satellites and observatories, the time accuracy of each probe must be less than 0.5 μs. In actual ground tests, the relative time accuracy of the double satellite is less than 0.12 μs. The test results at other times will be introduced in a special article. (Xiao et al submitted)

## 4.2 SiPM Event Acquisition Results

Considering signal pile-up and baseline instability at high counting rates, the event data at different counting rates were collected using Cd109 and Am241 sources. The fitted energy resolution at 88 keV increases little from 18 kcps to 132 kcps, and the sigma increases from 49 ADC to 53 ADC, indicating high antistacking performance for the proposed real-time data acquisition algorithm (Figure 7).

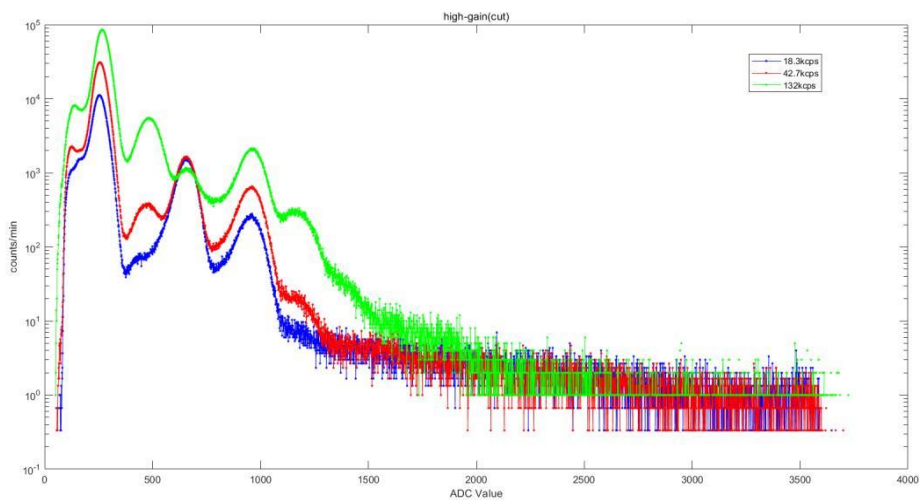

Figure 7 Energy Spectrum for Different Counting Rates

## 4.3 Maximum Event Transmission Capacity

The FIFO cache is used for 12-channel event packets to prevent signal loss from a specific channel being blocked by the sender. As the download frequency is 8 MHz, except for engineering parameters and time data packets, the maximum transmission event packet for each data acquisition board is set to be approximately 143 k. Each FIFO is sent in turn by polling to prevent congestion from channels with high event rates. When the maximum transmission event rate is exceeded, random data loss occurs in the channels with high event rates. The quantity of lost data can be queried through the flux.

On July 3, 2021, a major solar flare measuring X1.5 at its peak erupted. The maximum counting rate of many GRDs was almost 200 kcps, which lasted for more than 10 minutes. The total counts of every DAQ board exceeded the maximum transmission event rate. Figure 8 shows the changes between the total processing event rate and the total transmission event rate of the DAQ boards. The first spike in the middle panel of Figure 8 shows the buffering effect of using the FIFO cache. Due to the long-term high count rate, the FIFO cache was quickly saturated. The DAQ board maintains the maximum transmission rate until the processing rate is lower than the maximum. The bottom panel shows that the DAQ boards began to lose events randomly so that the complete loss of events in a channel did not occur.

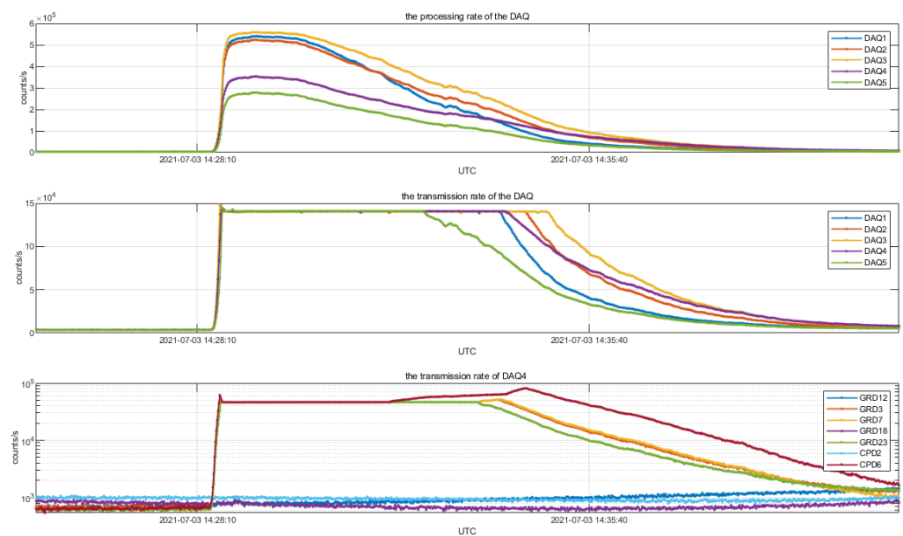

Figure 8 Maximum transmission event rate test results. Different colored lines represent different DAQ boards or detectors. The upper panel displays the processing event rate of each DAQ. The middle panel shows the transmission event rate of these DAQs. The bottom panel shows transmission rate of each detector of the DAQ board with the highest total count rate.

# 5 Conclusion

The design and implementation of a real-time signal acquisition algorithm for use in GECAM satellites is presented in this paper. The detectors of a satellite are classified into 5 groups, each of which independently acquires signals in parallel. The detector signal waveforms are digitized, and events are discriminated, processed, and stored by the FPGA. The test results show a high antistacking performance for the algorithm, even at an incidence of approximately 133 kcps. The caching method is optimized under transmission constraints. In the case of an extremely high event rate and saturation due to transmission bandwidth constraints, event packets are randomly lost to prevent a specific channel from being blocked by the sender. The algorithm has excellent performance when processing under the engineering time requirements and resource constraints. Many digital waveform processing methods, such as the sliding average methods, are worth exploring further.

**Acknowledgments**

The authors would like to thank all colleagues for helpful suggestions and comments. This study was supported by the National Natural Science Foundation of China (Grant No. 11803039 and 12173038) and the "Strategic Priority Research Program" of the Chinese Academy of Sciences (Grant Nos. XDA 15360100 and XDA 15360102).